\documentclass{article}

     \PassOptionsToPackage{numbers, compress}{natbib}

     \usepackage[preprint]{ml4molecules_2022}

\usepackage[utf8]{inputenc} 
\usepackage[T1]{fontenc}    
\usepackage{hyperref}       
\usepackage{url}            
\usepackage{booktabs}       
\usepackage{amsfonts}       
\usepackage{nicefrac}       
\usepackage{microtype}      
\usepackage{xcolor}         
\usepackage[normalem]{ulem} 
\usepackage{graphicx}       
\usepackage[rightcaption]{sidecap}

\title{Re-evaluating sample efficiency in de novo molecule generation}

\author{%
  Morgan Thomas \\
  Centre for Molecular Informatics\\
  University of Cambridge\\
  Cambridge, UK, CB2 1EW\\
  \texttt{mct50@cam.ac.uk}\\
  \And
  Noel M. O'Boyle \\
  Sosei Heptares \\
  Cambridge, UK, CB21 6DG\\
  \texttt{noel.o'boyle@soseiheptares.com}
  \And
  Andreas Bender \\
  Centre for Molecular Informatics\\
  University of Cambridge\\
  Cambridge, UK, CB2 1EW\\
  \texttt{ab454@cam.ac.uk}\\
  \And
  Chris de Graaf \\
  Sosei Heptares \\
  Cambridge, UK, CB21 6DG\\
  \texttt{chris.degraaf@soseiheptares.com}\\
}

\begin{document}

\maketitle

\begin{abstract}
\emph{De novo} molecule generation can suffer from data inefficiency; requiring large amounts of training data or many sampled data points to conduct objective optimization. 
The latter is a particular disadvantage when combining deep generative models with computationally expensive molecule scoring functions (a.k.a. oracles) commonly used in computer-aided drug design. Recent works have therefore focused on methods to improve sample efficiency in the context of \emph{de novo} molecule drug design, or to benchmark it. In this work, we discuss and adapt a recent sample efficiency benchmark to better reflect realistic goals also with respect to the quality of chemistry generated, which must always be considered in the context of small-molecule drug design; we then re-evaluate all benchmarked generative models. We find that accounting for molecular weight and LogP with respect to the training data, and the diversity of chemistry proposed, re-orders the ranking of generative models. In addition, we benchmark a recently proposed method to improve sample efficiency (Augmented Hill-Climb) and found it ranked top when considering both sample efficiency \emph{and} chemistry of molecules generated. Continual improvements in sample efficiency and chemical desirability enable more routine integration of computationally expensive scoring functions on a more realistic timescale. 
\end{abstract}

\section{Introduction}\label{intro}

Many deep generative model approaches applied to \emph{de novo} molecule generation for drug design have been proposed, most of which optimize molecule generation towards a particular endpoint using techniques such as reinforcement learning (RL) \cite{Olivecrona2017, Popova2018, Zhou2019} or Bayesian optimization \cite{Gomez-Bombarelli2018, Garcia-Ortegon2022}. In particular, reinforcement learning can be very sample inefficient, requiring up to $10^{5}$ samples to optimize towards an objective \cite{Brown2019, Olivecrona2017}. This serves as a practical limitation when using computationally expensive oracles to evaluate \emph{de novo} molecule fitness used to guide or update a generative model. This restricts even the use of routine computer-aided drug design techniques such as docking. Moreover, the use of docking as an oracle results in more diverse, bioactive-like chemistry compared to an ML-based oracle \cite{Thomas2021a}, and can be utilized when ligand data is scarce.

Several works have tried to improve sample efficiency. These include data augmentations during optimization \cite{Olivecrona2017, Korshunova2021}, reshaping of complex objectives into a series of smaller sequential tasks \cite{Goel2021, Guo2022a}, dynamic hyperparameter control \cite{Patronov2021} or algorithmic changes \cite{Fialkova2021, Atance2021, Thomas2022}. In particular, Thomas et al. \cite{Thomas2022} recently proposed Augmented Hill-Climb with recurrent neural networks evidencing significant sample efficiency improvement (on average 40-fold over a variety of thresholds) over the baseline REINVENT when using docking oracles. Additionally, the algorithm was applied to a GPT-style \cite{Radford2018} autoregressive architecture evidencing similar performance benefits. The authors compared this approach to other RL algorithms with recurrent neural networks, however, it is currently unknown how this compares to alternative generative model architectures and optimization algorithms.

Several benchmarks have been proposed for the comparison of generative model approaches for \emph{de novo} molecule generation \cite{Brown2019, Polykovskiy2020, Cieplinski2020, Garcia-Ortegon2022}, however, only recently has one been proposed that focuses on sample efficiency \cite{Gao2022}. Gao et al. \cite{Gao2022} proposed a practical molecular optimization benchmark (PMO) and compared a wider range of generative models on a selection of commonly used objective tasks. The authors reformulate performance as maximizing an objective within a fixed budget (10,000 oracle evaluations) and measure performance by comparing the area under the curve (AUC) of the average top 10 molecules during optimization. It was found that REINVENT \cite{Olivecrona2017} was the most sample efficient of 25 generative models implemented across all 23 tasks. 

\paragraph{This work} Here, we critically assess this benchmark and highlight key issues with respect to the \emph{de novo} chemistry proposed. We re-implement this benchmark and re-evaluate the results to better reflect the chemistry proposed by each model, by accounting for simple property filters and molecule diversity. Given the recent proposal of Augmented Hill-Climb, we also add this approach and assess performance compared to a wider variety of models than in the original publication \cite{Thomas2022}.

\section{Methods}\label{methods}

We re-implement the PMO benchmark using the code and data provided by the authors \cite{Gao2022} under an MIT license with no additional changes besides the implementation of Augmented Hill-Climb and modification of performance metrics as outlined below.

\subsection{Implementation of Augmented Hill-Cimb}\label{AHC_implementation}
We followed the recurrent neural network architecture of REINVENT \cite{Olivecrona2017} as implemented in the PMO benchmark. More specifically, the architecture consists of an embedding layer of size 128 and 3 layers of Gated Recurrent Units \cite{Cho2014} with a size of 512. The prior was trained on the PMO benchmark dataset (ZINC250k a subset of ZINC15 \cite{Sterling2015}) provided as is, using SMILES notation, and a batch size of 128 for a total of 5 epochs. For agent optimization, we use the Augmented Hill-Climb RL strategy outlined here \cite{Thomas2022}. A patience of 5 was used and the hyperparameters were optimized as in the PMO benchmark (Fig. \ref{fig:S1}) resulting in the following settings: $batch\_size=256, \sigma=120, K=0.25$. However, we additionally benchmarked $\sigma=60$ referred to as SMILES-AHC*. Note we omit any diversity filter or non-unique penalization to conduct a standardized comparison despite it being shown to improve performance \cite{Thomas2022}.

\subsection{Benchmark metric modifications}\label{benchmarkmods}
Measurement by the AUC of the ten highest-ranking molecules generated during optimization is retained to account for sample efficiency. However, we modify how these molecules are identified. 
\paragraph{AUC Top-10 (Filtered)} To include a property constraint to the initial training data, this metric filters out molecules that have a molecular weight or LogP beyond 4 standard deviations from the mean of pre-training dataset ZINC250k ($\mu\pm4\sigma$ approximately contains 99.99\% of a normal distribution). As a measure for topological idiosyncrasies, we additionally filter out molecules that contain more than 10\% \emph{de novo} (unobserved in ZINC250k) ECFP4 \cite{Rogers2010} fingerprint bits as proposed here \cite{Walters} and implemented in RDKit \cite{RDKit}. The 10\% threshold was chosen by inspecting molecules with varying fractions of \emph{de novo} ECFP4 bits and could therefore be refined in future work. We are of the opinion that these simple and lenient filters should be satisfied as a minimum requirement to ensure that the generative model does not drift beyond its applicability domain (if the model is distribution-based), or at-least maintains \emph{some} similarity to the training dataset on the basis that the dataset contains practically relevant chemistry. 
\paragraph{AUC Top-10 (Diverse)}
Gao et al. rationalise the selection of the highest-ranking ten molecules as "distinct molecular candidates to progress to later stages of development". We explicitly enforce this by selecting ten diverse molecules iteratively, where a molecule is only added to the selection if its Tanimoto similarity to any previously selected compounds is not higher than 0.35 (by ECFP4 fingerprints as implemented in RDKit). We use this threshold as anything more similar broadly correlates to an 80-85\% probability of belonging to the same bioactivity class \cite{Sayle2019, Jasial2016}, but distinct candidates should ideally possess different profiles.
\paragraph{AUC Top-10 (Combined)} A combination of applying both property filters and diversity filters as described above.

\section{Results \& Discussion}

\subsection{A critical assessment of the practical molecular optimization benchmark}
\begin{SCfigure}[0.9][h]
  \centering
  \includegraphics[width=0.5\textwidth]{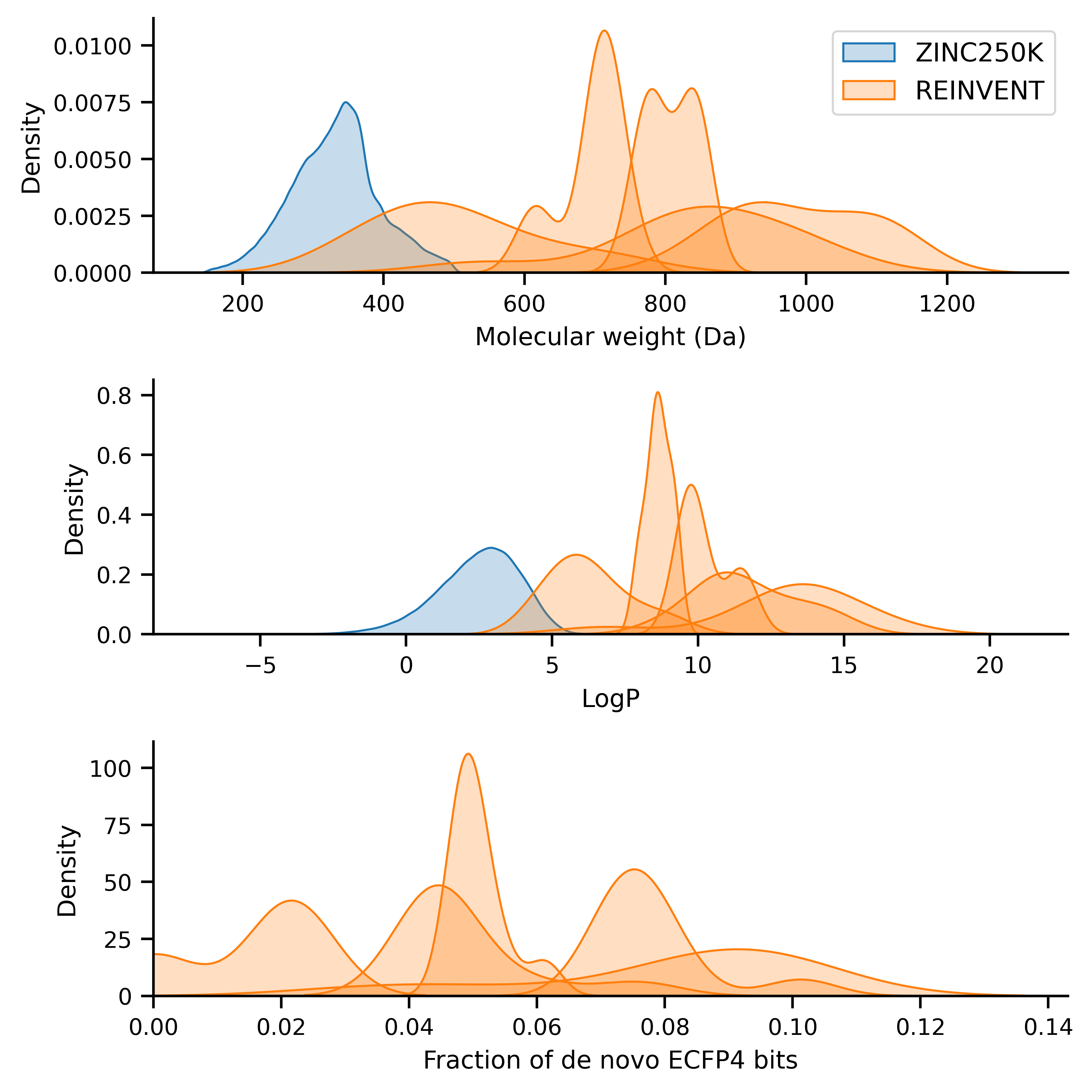}
  \caption{Properties of the top 10 \emph{de novo} molecules proposed by the 5 replicates of REINVENT optimization on the JNK3 task compared to the initial training dataset ZINC250k.}
  \label{fig1:sub1}
\end{SCfigure}

\begin{figure}[h]
  \centering
  \includegraphics[width=\textwidth]{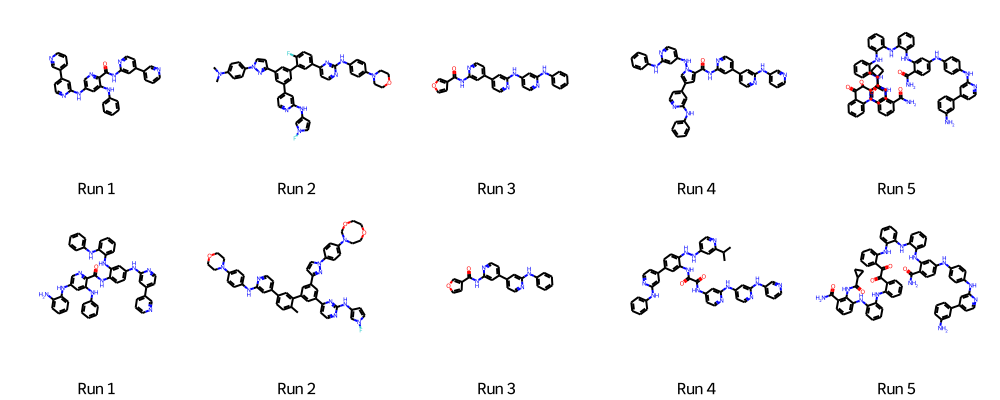}
  \caption{Top 2 molecules proposed by each of the 5 replicates of REINVENT optimization on the JNK3 task.}
  \label{fig1:sub2}
\end{figure}

We first sought to evaluate the top-performing model on the benchmark, REINVENT \cite{Olivecrona2017}. Before running the benchmark, the authors conducted hyperparameter optimization on two objective tasks which lead to selecting $\sigma=500$, with $\sigma$ being the reward coefficient that controls the balance between prior regularization or reward focus. Higher values of $\sigma$ decrease the loss contribution from the prior which decreases the regularization of the agent by the prior. In other work, it has been shown that values of $\sigma=240$ are high enough to lead to extrapolation outside the property space of the initial training dataset \cite{Thomas2022}. Therefore, we selected an objective where REINVENT particularly outperformed other generative models (JNK3) and investigated the property space and topology of the ten highest-ranking molecules proposed by REINVENT for each replicate run, and compared them to the initial ZINC250k pre-training dataset (shown in Fig. \ref{fig1:sub1} \& \ref{fig1:sub2}). Fig. \ref{fig1:sub1} shows that 4 of the 5 replicates result in the selected molecules having distributions of molecular weight and LogP far higher than ZINC250k. To reflect topological differences, we measured the ratio of \emph{de novo} fingerprint bits, showing that the selected molecules contained $0-10\%$ bits that were unobserved in ZINC250k. Alternatively, visualizing the top 2 for each run in Fig. \ref{fig1:sub2} highlights the undesirability from a chemical perspective, with large molecules and many repeating substructures. It is clear that in its current form the benchmark is limited to sample efficiency \emph{only}, and the results should be interpreted with caution. To serve as a practical comparison between generative models, it must reflect other desirable properties of \emph{de novo} molecule generation. 


\newpage
\subsection{Re-evaluating sample efficiency}
\begin{figure}[h]
    \centering
    \includegraphics[width=\textwidth]{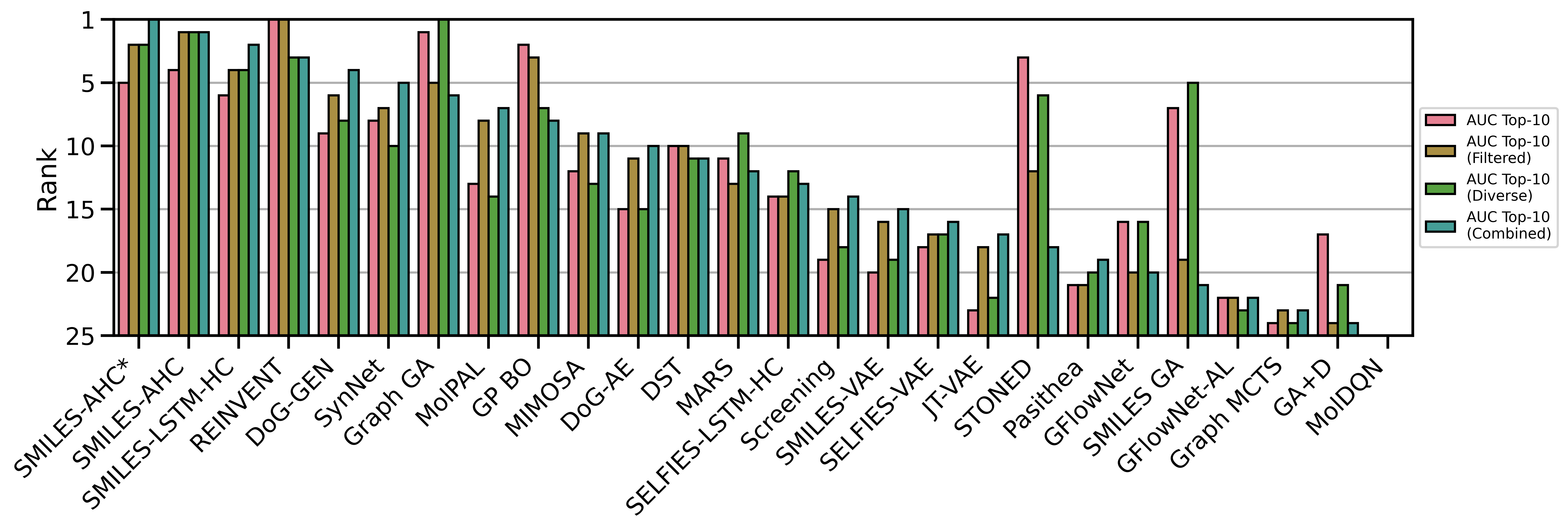}
    \caption{Rank of generative models according to each AUC Top-10 metric.}
    \label{fig2:sub1}
\end{figure}

\begin{figure}[h]
    \centering
    \includegraphics[width=\textwidth]{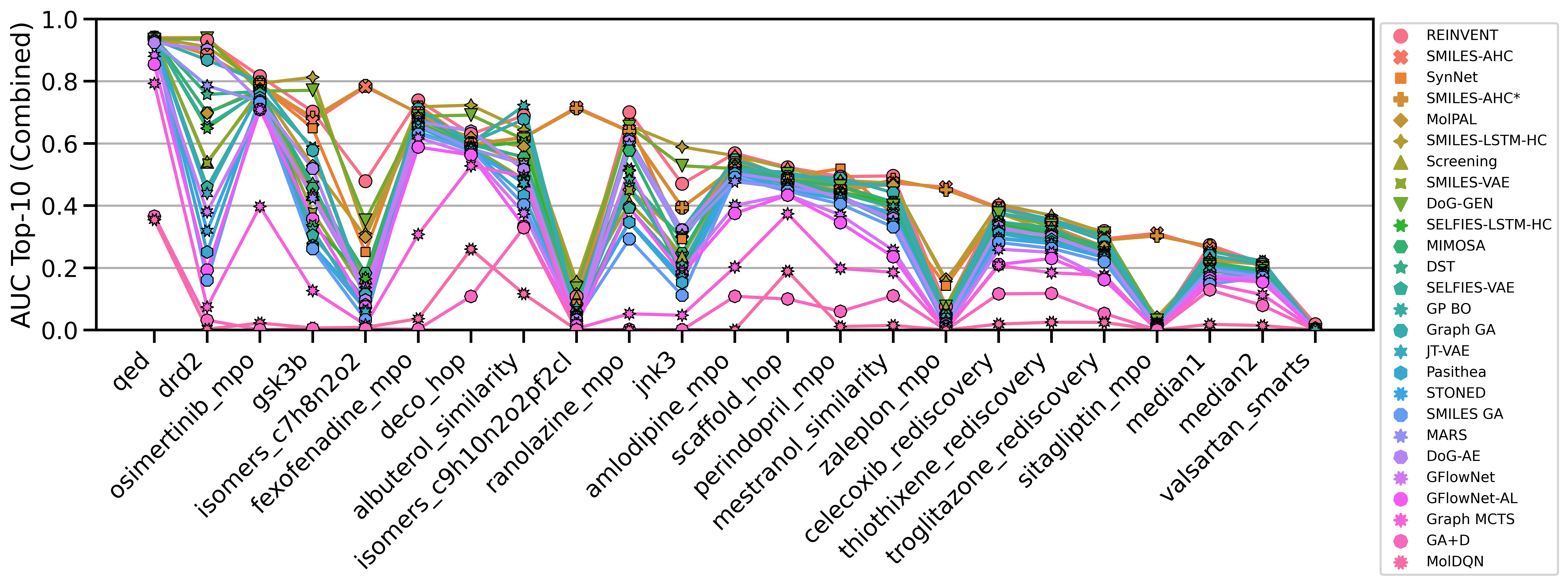}
    \caption{AUC Top-10 (Combined) for each generative model and objective task.}
    \label{fig2:sub2}
\end{figure}

Due to the lack of consideration for chemistry and examples of chemistry we observed, we designed new metrics to re-evaluate all generative models in the PMO benchmark (see \ref{benchmarkmods}) and further added Augmented Hill-Climb to the benchmark \cite{Thomas2022}. Fig. \ref{fig2:sub1} shows how the rank of generative models changes with the proposed AUC Top-10 metrics. Notably, it is SMILES-AHC* - where $\sigma$ was chosen based on prior knowledge of chemistry generated \cite{Thomas2022} - that achieves state-of-the-art performance when considering both property filters and diversity combined. Without imparting prior knowledge, the data-driven hyperparameter optimization of SMILES-AHC would rank first when accounting for property filters, diversity or both. Somewhat surprisingly, REINVENT still ranks first even with the property filters, suggesting that the minority of compounds that do not drift into undesirable property space still perform well. Meanwhile, other considerable re-rankings occur, especially for evolutionary-based algorithms such as Graph GA \cite{Jensen2019}, GP BO \cite{Garcia-Ortegon2022}, and SMILES GA \cite{Brown2019} that drop in rank when applying new metrics (unsurprising as these are rule-based as opposed to learning the ZINC250k distribution). Fig. \ref{fig2:sub2} shows the performance for each oracle. Interestingly, both SMILES-AHC and SMILES-AHC* achieved state-of-the-art by performing markedly better on the empirically more difficult tasks, such as both isomer-based tasks, Zaleplon MPO and Sitagliptin MPO. Overall, this ranking is more reflective of practical usage for generative models in drug design. 


\newpage
\section{Limitations \& Future work}\label{limitations}
 It should be noted that the results presented here are preliminary work, as all generative models have not undergone hyperparameter optimization against the new performance metrics proposed here. We leave this for future work. 
 
 Property filter thresholds can be highly subjective, hence, here we pick more lenient thresholds as a minimum requirement. The property filters and diversity requirements proposed here could alternatively be explicitly incorporated into the objective function as further oracles in a multi-parameter optimization setting. However, doing so is not a guarantee of the same results as shown here, as some models may better handle multiple parameters than others. Regardless of explicit incorporation, they should always be considered during evaluation, otherwise, benchmark results and hyperparameter settings may be misleading to non-domain experts.

Rule-based generative models are not constrained by the proposed standardized dataset of ZINC250k, which may only serve as an initial starting population. Therefore, comparing this to distribution-based models and measuring performance based on property similarities to ZINC250k will likely result in a bias toward distribution-based models. However, measuring desirable 'drug-like' chemistry objectively is a far more difficult task than comparison to a representative 'drug-like' dataset. The bias towards distribution-based models could be addressed by explicitly incorporating filters into the objective, or by instead using a larger dataset than ZINC250k that encompasses as much desirable 'drug-like' space as best as possible.

Most of the objective tasks used in the PMO benchmark derive from the GuacaMol \cite{Brown2019} benchmark, although GuacaMol uses a broader pre-training dataset derived from ChEMBL \cite{Mendez2019b}. As such, we investigated the compatibility between the new proposed ZINC250k dataset and the objective tasks. Fig. \ref{figS2:sub1} \& \ref{figS2:sub2} show that six objective task reference molecules sit in the lowest 0.01\% of molecular weight and LogP property space of ZINC250k. This raises questions as to whether it is reasonable or not to expect distribution-based generative models to be able to optimize for these objectives. Or, perhaps whether this is just a difficult challenge to find solutions in such an under-represented property space of the pre-training dataset.

\bibliography{main}

\appendix
\newpage
\section{Appendix}
\begin{figure}[h]
    \centering
    \includegraphics[width=\textwidth]{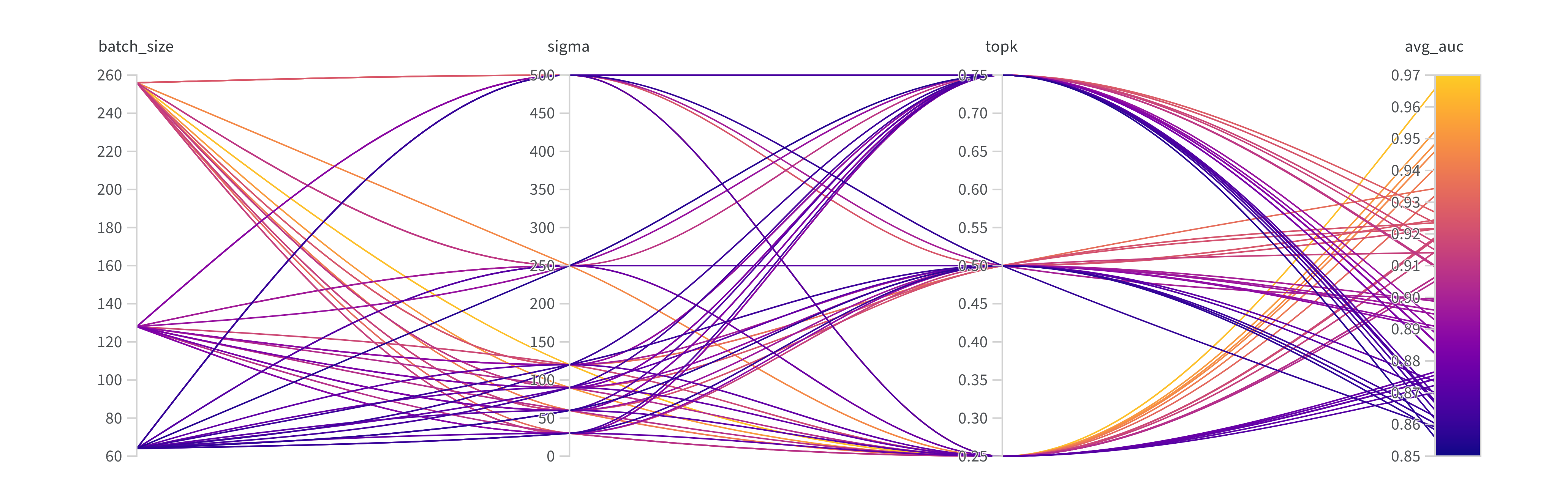}
    \caption{Hyper-parameter optimization of Augmented Hill-Climb on the two test objectives Zaleplon MPO and Perindopril MPO.}
    \label{fig:S1}
\end{figure}


\begin{SCfigure}[0.9][h]
  \includegraphics[width=0.5\textwidth]{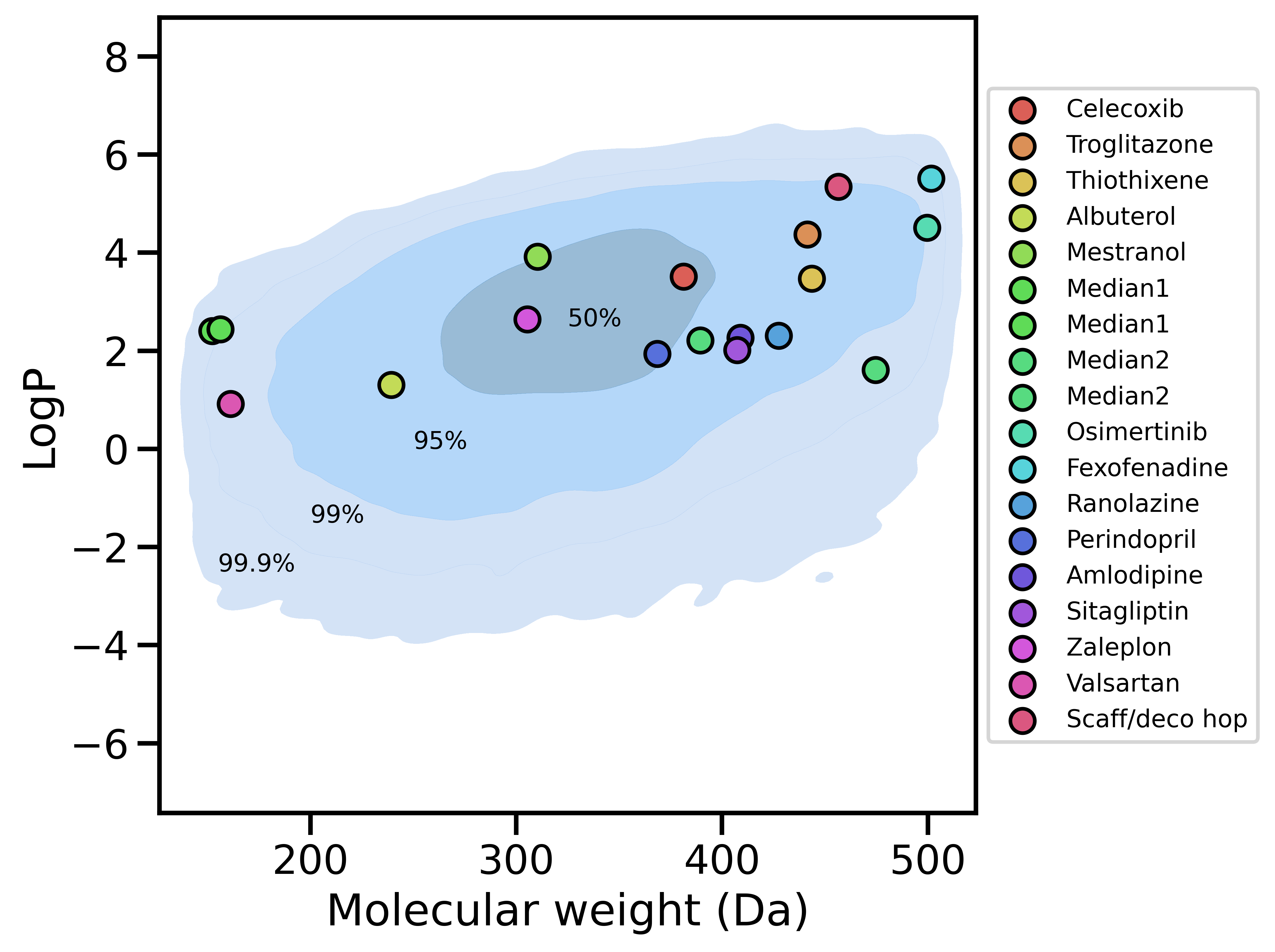}
  \caption{Property space of benchmark task reference molecules in comparison to ZINC250k (up to 99.99\% probability mass), no molecules have more than 10\% \emph{de novo} ECFP4 bits.}
  \label{figS2:sub1}
\end{SCfigure}

\begin{SCfigure}[0.9][h]
  \includegraphics[width=0.5\textwidth]{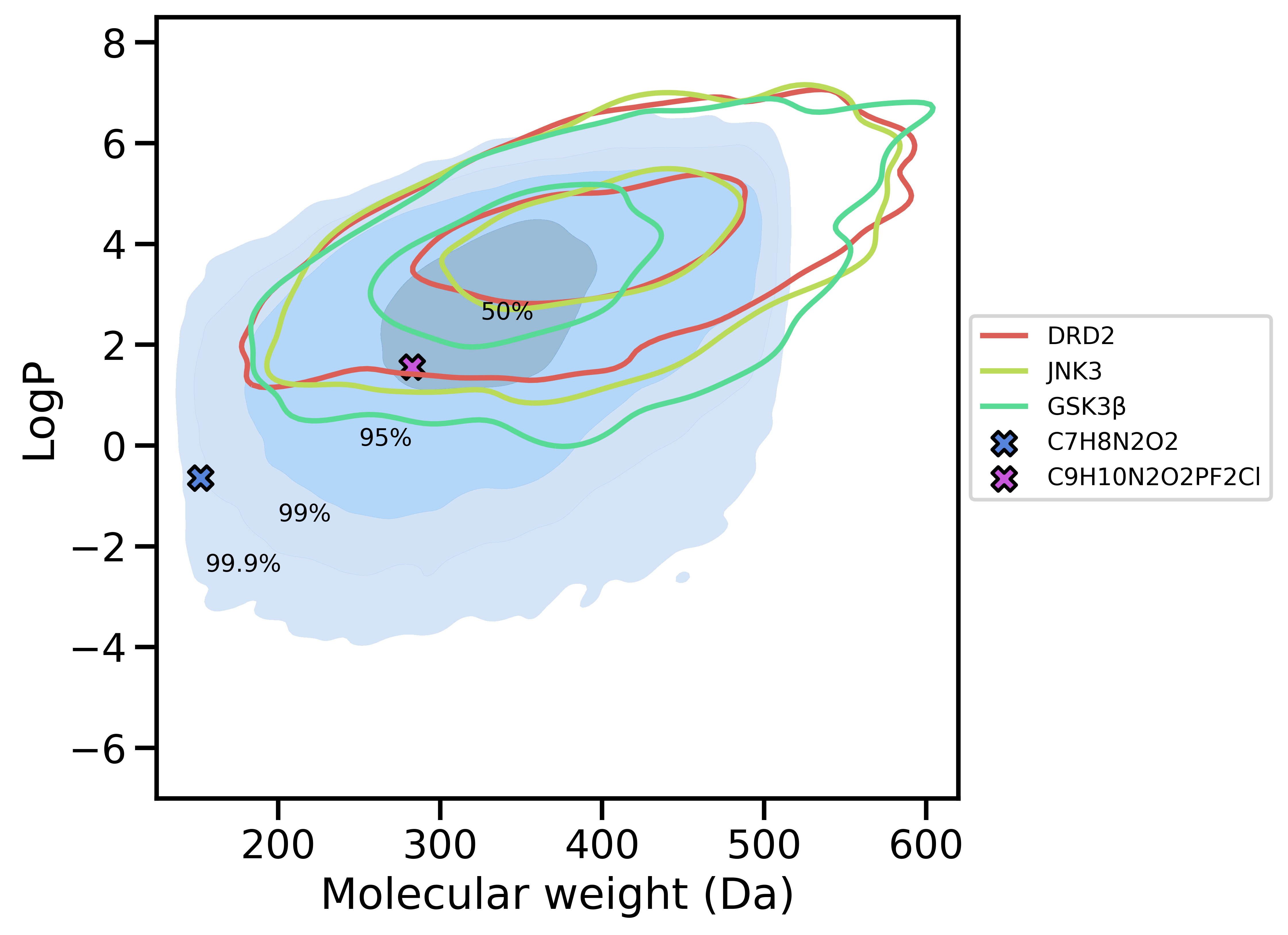}
  \caption{Property space of benchmark task reference distributions in comparison to ZINC250k. Isomers were generated using MAYGEN \cite{Yirik2021} and the mean (of a random sample of 1000) respective property plotted. All generated isomers had more than 10\% \emph{de novo} ECFP4 bits. DRD2, JNK3 and GSK3$\beta$ active molecules were downloaded from ExCAPE-DB \cite{Sun2017} and their distribution plotted up 95\% probability mass. DRD2, JNK3 and GSK3$\beta$ had 13.4\%, 7.7\% and 16.9\% of molecules with more than 10\% \emph{de novo} ECFP4 bits respectively.}
  \label{figS2:sub2}
\end{SCfigure}

\end{document}